# Comparative study of pyroelectric response of PZT-film/Si, PZT-film/Por-Si/Si and PVDF-film/Si structures

S. Bravina[a], N. Morozovsky[a], E.A. Eliseev[b], E. Cattan[c], D. Remiens[c] and A. Grosman[d]

[a]Institute of Physics of NAS of Ukrainre, 46 Prospect Nauki, 03028 Kiev, Ukraine
E-mail: bravina@iop.kiev.ua

[b]Institute for Problems of Materials Science, National Academy of Science of Ukraine, Krjijanovskogo 3, 03142 Kiev, Ukraine,
E-mail: eliseev@mail.i.com.ua

[c]OAE-dept. IEMN, UMR 8520 Universite de Valenciennes et du Hainaut-Cambresis, ZI de la petite savate, 59600 Maubeuge, France

[d]Groupe de Physique des Solides, UMR 7588, Universites Paris 6&7, 2 Place Jussieu, 75251 Paris Cedex 05, France

The comparative investigation of pyroelectric response modulation frequency dependences of structures of "polar active film-Si substrate" type based on PZT and PVDF and of structures of "polar active film - buffer layer - Si substrate" type based on PZT and porous silicon (por-Si) has been carried out. By photopyroelectric modulation method the thermowave pyroelectric "under-surface" probing was performed and amplitude-to-frequency and phase-to-frequency dependences of pyroelectric response of investigated systems in the voltage and current modes were obtained.

By performing the analysis of obtained dependences the thermal diffusivity values of investigated PZT and PVDF films, and also por-Si interlayer were estimated. The results of theoretical consideration of the structures under investigation are in a good agreement with the experimental data.

The problem of thermal decoupling and self-decoupling from position of chain "frequency-thickness-material" and question of complete and incomplete thermal linkage are briefly discussed. The obtained results show that a por-Si layer is a suitable material for effective thermal decoupling of polar active film and heat removing Si-substrate which realized in significant increase of pyroelectric response value and approach it to that of characteristic for a free sensitive element.

## 1. Introduction

Ceramics of Pb(Zr$_x$Ti$_{1-x}$)O$_3$, (PZT) and polymer polyvinilydenefluoride (-CH$_2$-CF$_2$-)$_n$, n≥10$^4$, (PVDF) are well-known polar-active materials [1 - 7].

Application of PZT ceramics started in the sixties and very quickly spread firstly into the area of piezoelectric transducers [5] then for plane and cavity pyroelectric converters [3, 6] and now is spreading into the area of micro-and electromechanics and memory elements [2, 7].

Application of PVDF, with temperature dependent dipolar moment higher in comparison with the other polymers, has started since the seventies [1, 3, 4]. At present on the base of PVDF are manufactured piezo- and pyroelectric converters for various purposes and of various configurations [3, 6] in both plane and cavity design [8].

Ceramics of PZT type are obtained by high temperature synthesis and crystallization of furnace charge of definite composition (for PZT ceramics Zr/Ti ratio defines the temperatures of phase transition and vicinity to morphotropic boundary) [5]. Piezoelectric and pyroelectric properties of PZT ceramics are acquired as the result of high temperature electric-field treatment.

Thick films of PVDF acquire polar properties in the course of mechano-thermo-electrical treatment (stretching in ratio from 1:4 to 1:20, polarization in electrical field ~100 MV/m under temperature $T_t$ = 80-120 $^o$C) results in formation and fixing of polar β-phase [4, 9].

Polar properties of thin (0.01-1mm) PZT plates and thick (10-100μm) PVDF films are remained also in thinner (≤1μm) films of these materials [6, 9]. Film technology of both PZT and PVDF in principal is compatible with Si-basis of modern microelectronics. In particular, piezo-and pyroactive structures of "polar active film on Si-substrate" type based on PZT films can be obtained by sol-gel [10] and RF sputtering techniques [11] and those based on PVDF films are manufactured by vacuum deposition [12] and by sol-gel spin coating [13].

PZT films deposited on Si-substrate acquire polar properties after temperature treatment at $T_t$ = 550-650 $^o$C due to a stable perovskite phase formation, which is achieved under cooling to room temperature. To acquire piezo and pyroelectric properties of PVDF films on Si-substrate the structures are subjected to electric-field treatment at $T_t$=80-120 $^o$C.

The advantage of PZT films over PVDF films are high values of pyroelectric coefficient (≈3·10$^{-4}$ C/m$^2$K) and the temperature of reversible retention of polar activity (≈200$^o$C), but rather high value of dielectric permittivity (from ≈300 to ≈1000 for different

Zr/Ti ratios)). The advantage of PVDF films are rather low values of dielectric permittivity ($\approx 11$) and thermal diffusivity ($\sim 10^{-7}$ m$^2$/c) and also high mechanical flexibility and low fabrication costs but rather low value of pyroelectric coefficient ($\approx 3 \cdot 10^{-5}$ C/m$^2$K) and the temperature of keeping polar activity ($\approx 60$ °C).

The both considered polar active materials are the basis of sensitive elements (SE) of most modern Si-integrated pyroelectric converters including pyroelectric detectors (PyED). It is known that efficiency of conversion of PyED is determined not only by properties of SE and substrate material but also can be controlled by using of intermediate buffer layer. This layer with high (heat sink) or low (heat insulator) value of thermal conductivity, which placed between sensitive layer and its substrate in case of low value of thermal conductivity plays role of thermal coupling or decoupling between sensitive pyroactive layer and its substrate. At present, the well-known Si–derived materials with thermal conductivity reduced in comparison with crystalline Si, namely SiO$_x$, SiO$_2$-Si$_3$N$_4$ systems, amorphous Si and porous Si are considered as promising for heat isolating layers. Such a layer can be used for buffer layers of SE of PyED to improve their sensitivity.

We performed the comparative investigation of pyroelectric response frequency dependences of structures of "polar active film-Si substrate" type based on PZT and PVDF and of structures of "polar active film - buffer layer - Si substrate" type based on PZT and porous silicon (por-Si).

## 2. Structure preparation

Investigated Pt-PZT-Pt/Ti-SiO$_2$/Si structures with oriented PZT layer were manufactured by RF magnetron sputtering in the system and under conditions described previously [11].

The sputtering target obtained by uniaxial cold pressing includes the mixture of PbO, TiO$_2$ and ZrO$_2$ in a stoichiometric composition. The structure includes the top 150 nm Pt-electrode, 1.9 μm layer, bottom Pt/Ti-bilayer (150 nm of Pt, 10 nm of Ti) deposited onto the oxidized (350 nm of SiO$_2$) (100) n-type Si 350 μm substrate.

For PZT – Si-substrate structure is necessary to design the bottom electrode, which possesses not only a stable and high enough electrical conductivity but also simultaneously prevents the interfacial reactions between electrode, PZT and SiO$_2$ components in PZT-film and Si-substrate surroundings under rather high temperatures [14]. The thickness of Ti-layer plays an important role in limiting the diffusion of Ti in Pt/Ti intermediate bilayer through Pt-layer into the PZT-layer and directly into SiO$_2$-layer, and also in correction of poor adhesion of Pt-layer [15, 16]. In correspondence with the developed procedure [11, 16] we obtained the $TiO_x$-layer in which the processes of Ti-diffusion are significantly suppressed.

The annealing treatment of the $Pt/TiO_x/SiO_2/Si$-substrate structure just before of PZT deposition was performed for substrate stabilization and post-annealing treatment of PZT-film was performed for crystallizing obtained film in the polar perovskite phase.

The obtained PZT-films with Zr/Ti ratio 54/46 are near to the morphotropic phase boundary, which corresponds to best performances for bulk PZT ceramics. Deposited by sputtering procedure, which was followed by a lift-off, the top Pt- electrodes have 1 $mm^2$ of area. Each irradiated element has current-carrying thin stripe with circle current electrode of 1 mm of diameter on the stripe end.

Investigated Pt-PZT-Pt/Ti-porSi/$SiO_2$/Si structures with 1.4 µm PZT layer were manufactured by similar way. Each irradiated of 1x1 $mm^2$ of area element has current-carrying thin stripe with current electrode of 0.1 $mm^2$ of area on the stripe end.

Porous silicon layer is prepared by electrochemical etching of Si single crystal in a HF solution. The porosity (from ≈30% to ≈90%) is controlled by the current density and the HF concentration and the thickness of the layer (from ≈1µm to a few hundred micrometers) by the time of the anodic dissolution [17]. Porosity and thickness are determined with high precision by weighing. The porous layer under study was prepared from highly boron doped [100] Si substrate. The morphology of such layers can be described as honeycomb-like structure with pores perpendicular to the Si substrate [17, 18]. The pores of polygonal sections, with mesoscopic sizes, are separated from each other by Si single crystal walls of constant thickness of 10nm order. The electrochemical formation conditions were: EtOH:HF (1:1) solution, 50 mA/$cm^2$, 5 minutes. The porosity was 65% and the thickness 10.4 µm. It is noteworthy that the porous layer exhibits the resistivity of intrinsic Si even when the Si substrate is heavily doped [19].

Investigated Al/Ti-PVDF-Ti-$SiO_2$/Si structures include the polar PVDF layer were manufactured by spin coating method. The solution of PVDF precursor in ethyl-methyl-ketone 10/90 (% w/w) at 50 °C with agitation during 1 h was prepared. Spin coating of filtered solution-gel on the Si-substrate coated by evaporated bottom electrode during 1 min was performed at 1000 rpm.

Finally the structure includes 1 µm thick 5x5 mm top Al- electrode, 25 nm Ti sublayer, 4.3 µm layer of PVDF, bottom 1 µm Ti electrode deposited onto [100] boron doped 625 µm Si substrate.

For both out-gassing and adherence purposes as well as for increase of crystallinity of the PVDF the thermal treatment at 170 °C during 1.5 hour was performed.

Poling process includes step-by-step increase-decrease of applied DC voltage up to 200-400 V at room temperatures or elevated ones.

## 3. Experiment

The measurements of the pyroelectric response were performed by photopyroelectric modulation method in which the samples under investigation are irradiated by modulated IR flux $\Phi(t) = \Phi_1 + \Phi_0 \cdot \exp(j\omega_m t)$ and generated pyroelectric response amplitude $U_\pi$ and phase $\varphi_\pi$ are measured as the functions of angular $\omega_m$ or cyclic $f_m = \omega_m/2\pi$ modulation frequency.

In this case the choice of $f_m$ permits to change the effective temperature wave penetration depth $\lambda_t = (a_t/\pi f_m)^{1/2}$ ($a_t$ – is the thermal diffusivity) which gives the possibility to perform the under-surface thermal wave probing. So, the thermally excited region of the sample depending on $f_m$ value is the volume (low enough $f_m$) or near-surface layer (high enough $f_m$).

The source of temperature wave in our case is the surface of material under characterization, temperature variation of which is caused by absorbed part of the energy of the modulated irradiation flux.

The measurements were fulfilled by means of the measuring system described in Ref. [20] developed for complete ferroelectric diagnostics. This system allows one to investigate the amplitude-frequency $U_\pi(f_m)$ and phase-frequency $\varphi_\pi(f_m)$ dependences of pyroelectric response in the wide range of modulation frequency $f_m$ values, of 1 Hz to 100 kHz, under short circuit and open circuit conditions, which correspond to the pyroelectric current mode and pyroelectric voltage mode.

For the measurements the electrodes of the sample under investigation were connected to the matching stage with step-varying input impedance [21] (from high value ~10 GΩ at 20 Hz in the pyroelectric voltage mode to low one ~100 kΩ in the pyroelectric current mode).

## 4. Theoretical consideration

### *4.1. Free sensitive element*

Solving [3, 9] the set of differential equations which describe the temperature variation $\vartheta(t)$ and pyroelectric response $U_\pi$ for the free SE of PDR in the shape of plate with metal electrodes on its main polar surfaces illuminated by the modulated thermal flux $\Phi(t)=\Phi_1+\Phi_0\cdot\exp(j\omega_m t)$

$$\frac{d}{dt}[C_{th}\vartheta(t)] + \vartheta(t)/R_{th} = \alpha\,\Phi(t)\,A_0, \tag{1}$$

$$\frac{d}{dt}(C_s U_\pi) + U_\pi/R_s = \frac{d}{dt}[\gamma\,A_0\,\vartheta(t)], \tag{2}$$

gives for voltage response $U_\pi(\omega_m, t)$ the following expression:

$$U_\pi(\omega_m, t) = \frac{\alpha\gamma\Phi_0 A_0^2 \tau_{th}\tau_e\,j\omega_m\exp(j\omega_m t)}{C_{th}C_s(1+j\omega_m\tau_{th})(1+j\omega_m\tau_e)} \tag{3}$$

where $\alpha$ is the absorptivity, $\gamma = dP_s/dT$ is the pyroelectric coefficient, $A_0$ is the area of the irradiated receiving surface of SE, $\tau_{th} = R_{th}C_{th}$ is the thermal time constant, $R_{th}$ and $C_{th}$ are the thermal resistance and capacitance, $\tau_e = R_s C_s$ is the electrical time constant, $R_s$ and $C_s$ are the electrical resistance and capacitance of the SE.

Usually $\omega_m\tau_{th} \gg 1$ and for SE loaded on external circuit with electrical resistance $R_L$ and capacitance $C_L$ the value of $U_\pi$ is given by expression [3, 20, 21]

$$U_\pi(\omega_m) = (\gamma/c_t)\,\alpha\,\Phi_0\,A_0\,R_e/d\,\sqrt{1+(\omega_m R_e C_e)^2} \tag{4}$$

where $c_t$ is the volume heat capacity of SE material, $R_e^{-1} = R_L^{-1} + R_s^{-1}$ and $C_e = C_L + C_s$, $R_s = d/\sigma A_e$ and $C_s = \varepsilon\varepsilon_0 A_e/d$ are the electrical resistance and capacitance of the SE, $d$ is the thickness of the SE, $A_e$ is the electrode area, $\varepsilon$ and $\sigma$ are dielectric permittivity and conductivity of SE material respectively, and $\varepsilon_0 = 8{,}85\cdot10^{-12}$ F/m.

In the pyroelectric current mode $\omega_m R_e C_e \ll 1$ and so

$$U_\pi = U_{\pi 1} = (\gamma/c_t)\alpha\Phi_0 A_0 R_L/d \tag{5}$$

At that $\varphi_\pi = \varphi_{\pi 1} = \text{const}(\omega_m)$ and $U_{\pi 1}$ is in phase or in the opposite phase with thermal flux intensity $\Phi(t)$ depending on the polarization direction in the SE.

In the pyroelectric voltage mode $\omega_m R_e C_e \gg 1$ and so

$$U_\pi = U_{\pi 2} = (\gamma/c_t\varepsilon)\alpha\Phi_0 A_0/\varepsilon_0 i_e\omega_m \tag{6}$$

At that $\varphi_\pi = \varphi_{\pi 2} = \text{const}(\omega_m)$ and $U_{\pi 2}$ and thermal flux intensity $\Phi(t)$ have the phase shift equals $\pm\pi/2$, depending on the polarization direction in the SE, and so $\varphi_{\pi 2} - \varphi_{\pi 1} = \pi/2$.

It results from (4) that with increase of $\omega_m$ the $U_{\pi 1}$-mode (5) measured at $R_L \ll R_s$ turns to $U_{\pi 2}$-mode (6) and the transition frequency $\omega_{mt}$ is determined by relation $\omega_{mt} R_L C_e = 1$. This so called "frequency effect" leads to realization of only $U_{\pi 2}$ mode at frequencies above $\omega_{mt}$.

### 4.2. Film sensitive element on the interlayer on heat sink substrate

The case of operation of sensitive element on substrate was earlier considered in [22-24]. In the case of heat transfer through a multilayer system of i = 1, 2, …n layers the distribution of the temperature variation $\vartheta_i(x, t)$ is described by the system of one-dimensional heat diffusion equations for each of the individual layers

$$\frac{\partial}{\partial t} \vartheta_i(x, t) = \frac{k_{ti}}{c_{ti}} \cdot \frac{\partial^2}{\partial x^2} \vartheta_i(x, t), \qquad (7)$$

where $k_{ti}$ and $c_{ti}$ are the thermal conductivity and volume heat capacity of each of the layer. The alternating component of general solution of the system (7) $\vartheta_i(x, t) = \vartheta_i(x) \exp(j\omega_m t)$ can be written as

$$\vartheta_i(x) = A_i \exp[(x - x_{i-1})/L_{ti}] + B_i \exp[-(x - x_i)/l_{ti}] \qquad (8)$$

where $L_{ti} = (a_{ti}/j\omega)^{1/2}$, $a_{ti} = k_{ti}/c_{ti}$ is the thermal diffusivity, and $A_i$, $B_i$ are determined from the boundary conditions at the interfaces between different layers followed from the equality of the temperature values on the layer boundaries and the continuity of the heat current density $j_i(x) = -k_{ti}(\partial \vartheta_i / \partial x)$:

on the front irradiated face $x = x_0$ $\quad j_1(x_0) = \alpha \Phi_0 - g_0 \vartheta_1(x_0)$

on the other layer boundaries $\quad j_i(x_i) = j_{i+1}(x_i), \quad \vartheta_i(x_i) = \vartheta_{i+1}(x_i)$ $\qquad (9)$

and on the rear face $\quad j_n(x_n) = g_n \vartheta_n(x_n)$

where $g_0$ and $g_n$ are thermal losses conductivity.

Using the relation $I_\pi = \gamma A_0 [\partial \overline{\vartheta}(t)/\partial t]$, between pyroelectric current $I_\pi(t)$ and the speed of thickness averaged temperature change $\partial \overline{\vartheta}(t)/\partial t$ and the relation $U_\pi = (\gamma/\varepsilon\varepsilon_0) d \overline{\vartheta}(t)$ between pyroelectric voltage $U_\pi(t)$ and averaged temperature change $\overline{\vartheta}(t)$ with taking into account that $\overline{\vartheta}(t) = (1/d) \int_0^d \vartheta(x,t) dx$ gives the following expressions for $I_\pi(t)$ and $U_\pi(t)$:

$$I_\pi = (\gamma/d)A_0 \int_0^d \frac{\partial}{\partial t} \vartheta(x,t)dx \qquad (10)$$

$$U_\pi = (\gamma/\varepsilon\varepsilon_0) \int_0^d \vartheta(x,t)dx \qquad (11)$$

The solution of (7) under conditions (9) together with (10, 11) taking into account the relation $U_\pi(\omega_m)=I_\pi(\omega_m)\cdot R_e/[1+(\omega_m\tau_e)^2]^{1/2}$ between pyroelectric voltage $U_\pi(\omega_m)$ and pyroelectric current $I_\pi(\omega_m)$ for circuit with the structure of SE in the shape of the electroded plate gives the dependences $U_{\pi 1,2}(\omega_m)$ and $\varphi_{\pi 1,2}(\omega_m)$.

For three-layer system of SE - buffer layer - substrate the obtained analytical solution of (7)-(9) is rather cumbersome, that is why we do not write it down here, but for two-layer system SE - substrate it can be easily analyzed for different cases of ratio $\delta = b_{t1}/b_{t2}$ for effusivities $b_{ti} = (k_{ti}c_{ti})^{1/2}$ of SE and substrate.

Thus, in general case the expression for $\bar\vartheta_1(t)= \bar\vartheta_1 \exp(j\omega_m t)$ of SE for the system of SE with thickness $d$ and substrate with thickness $L_S$ for values of $g_0$ and $g_2$ negligible in real cases has the following view

$$\bar\vartheta_1 = \frac{\alpha\Phi_0 L_{t1}^2}{k_{t1}d}\left[1-\frac{\delta\sinh(L_S/L_{t2})}{\cosh(L_S/L_{t2})\sinh(d/L_{t1})+\delta\cosh(d/L_{t1})\sinh(L_S/L_{t2})}\right] \qquad (12)$$

from which taking into account (10,11) is clearly seen that in the case of high value of $\delta$ the values of $U_{\pi 1}$ and $U_{\pi 2}$ decrease significantly and their frequency dependences are changed too. In the case of small values of $\delta\ll 1$ (for example thick enough air gap buffer) and also for high values of modulation frequencies for any $\delta$ values the solution reduces to that for the free SE described by expressions (4 - 6) for $U_{\pi 1,2}(\omega_m)$.

The theoretical dependences obtained for considered free SE and for the structures of film SE on Si-substrate and SE on por-Si interlayer on Si-substrate which are based on PZT and PVDF are presented in Fig. 1. The parameters used for our calculations are presented in Table 1.

As it can be seen in Fig.1 (curves PZT, PVDF) the behavior of $U_{\pi 1,2}(\omega_m)$ and $\varphi_{\pi 1,2}(\omega_m)$ of a free SE is in accordance with expressions (4)-(6), namely above the certain frequency $\omega_{mt}=1/R_L C_e$ is observed the transition from pyroelectric current mode, $U_{\pi 1}(\omega_m) = $ const, $\varphi_{\pi 1}(\omega_m) = 0°$, to pyroelectric voltage mode, $U_{\pi 2}(\omega_m) \propto \omega_m^{-1}$, $\varphi_{\pi 1}(\omega_m) = 90°$ (frequency effect).

**Table 1. Thermophysical parameters of the components of considered «Polar film-Si-substrate» systems [3, 25-27]**

| Material parameter | PZT | PVDF | Si | SiO$_2$ | Por-Si$^*$ | Al | Pt | Pt/Ti | Ti |
|---|---|---|---|---|---|---|---|---|---|
| Thermal conductivity, W/mK | 1.08 - 1,4 | 0.13–0.14 | 141-150 | 1.2–1.3 | 0.2–0.5 | 237 | 71 | 45.6 | |
| Volume thermal capacity, $10^6$ J/m$^3$K | 2.7 – 2.8 | 2.1-2.4 | 1.6-1.64 | 1.65 – 1.68 | ≈1 | 2.43 | 2.83 | 2.71 | 2.67 |
| Thermal diffusivity, $10^{-6}$ m$^2$/s | 0.39 -0.5 | 0.056–0.061 | 88-91 | 0.7 – 0.79 | 0.2 – 0.5 | 98 | 25.1 | 16.8 | |
| Thermal effusivity, $10^3$ W·s$^{1/2}$/m$^2$K | 1.74 -1.8 | 0.52–0.6 | 15–15.6 | 1.4–1.47 | 0.45 –0.7 | 2.4 | 14.2 | 11.1 | |

* our data

The influence of substrate (Fig.1, curves PZT/SiO$_2$/Si) is manifested in the decrease of $U_{\pi 1}(\omega_m)$ and $U_{\pi 2}(\omega_m)$ values, and also in a significant change of view of $U_{\pi 1,2}(\omega_m)$ and $\varphi_{\pi 1,2}(\omega_m)$ dependences. The peculiarities of these dependences are the low-frequency and high-frequency breaks of $U_{\pi 1}(\omega_m)$ and $U_{\pi 2}(\omega_m)$ and corresponding regions of sharp change of $\varphi_{\pi 1}(\omega_m)$ and $\varphi_{\pi 2}(\omega_m)$. For $U_{\pi 1}(\omega_m)$ and $\varphi_{\pi 1}(\omega_m)$ high frequency peculiarities are masked with the frequency effect.

The position of low-frequency peculiarities $\omega_{m1}$ are determined by relation of $\lambda_t \approx L_S$ and $\omega_{m1} \approx 2a_{tS}/L_S^2$, where $a_{tS}$ and $L_S$ are thermal diffusivity and thickness of the substrate. There is a "complete thermal linkage" of SE and substrate below $\omega_{m1}$. The position of high-frequency peculiarities $\omega_{m2}$ is determined by relation $\lambda_t \approx d$ and $\omega_{m2} \approx 2a_t/d^2$, where $a_t$ and $d$ are the thermal diffusivity and thickness of SE. Above $\omega_{m2}$ thermal linkage of SE and substrate is insignificant. Between $\omega_{m1}$ and $\omega_{m2}$ is applied the relation $d < \lambda_t < L_S$ and there is an incomplete thermal linkage of SE and substrate.

The existence of intermediate thermal isolating layer (thermal buffer) (Fig.1, curves PZT/porSi/Si) leads to shift of low-frequency and high-frequency peculiarities of $U_{\pi 1,2}(\omega_m)$ and $\varphi_{\pi 1,2}(\omega_m)$ into the region of low frequencies. The decrease of thermal coupling of SE and substrate leads to the occurrence of low-frequency region of $U_{\pi 1}(\omega_m) \propto \omega_m$ and the regions of sharp change of $\varphi_{\pi 1}(\omega_m)$ corresponding to the ends of this low-frequency region. Thermal decoupling of SE and substrate is noticeable when $\lambda_t \leq d_b$ so for the frequencies higher than $\omega_{mb} \approx 2a_{tb}/d_b^2$, where $a_{tb}$ and $d_b$ are the thermal diffusivity and thickness of buffer layer. Since $\omega_{mb} < \omega_{m2}$ occurs the shift of the region of incomplete thermal coupling of SE and substrate into the region of low frequencies. Due to this the significant expansion of the regions of $U_{\pi1,2}(\omega_m)$ and $\varphi_{\pi1,2}(\omega_m)$ characteristic for a free SE with simultaneous increase of pyroelectric response value is achieved.

It should be noticed similarity of view of $U_{\pi1,2}(\omega_m)$ and $\varphi_{\pi1,2}(\omega_m)$ calculated for PZT-film – por-Si interlayer – Si-substrate, (Fig. 1, curves PZT/porSi/Si) and PVDF-film – Si-substrate (Fig. 1, curves PVDF/SiO$_2$/Si) connected evidently with comparable high values of the linear thermal resistance $r_{th} = (1/k_t)d$ of investigated por-Si layer and PVDF film.

## 5. Results and coments

### 5.1. Free PZT and PVDF sensitive elements

The results obtained for the free SE based on PZT ceramic thin plate and for PVDF film are presented in Fig. 2a and Fig. 2b.

The behaviour of $U_{\pi1,2}(f_m)$ and $\varphi_{\pi1,2}(f_m)$ at low frequencies and $U_{\pi2}(f_m)$ and $\varphi_{\pi2}(f_m)$ at high frequencies corresponds to (5) and (6) which is the evidence of the lack of thermal and pyroelectric inhomogeneity. The difference of the observed $U_{\pi1,2}$ values of investigated PZT ceramic and PVDF film samples is connected with differences of their pyroelectric figures of merit $M_1 = \gamma/c$ and $M_2 = \gamma/(c\,\varepsilon)$ values [3], their geometries and also absorptivities of illuminated electrodes.

The change of the sign of poling voltage leads to 180 °- change of $\varphi_{\pi1,2}$ (Fig. 2a) which reflects the change of direction of pyroelectric reaction connected with the reverse of polarization direction.

The high frequency drop of $U_{\pi1}(f_m)$ and rise of $\varphi_{\pi1}(f_m)$ and also the gradual approach of $U_{\pi1}(f_m)$ to $U_{\pi2}(f_m)$ as well as the approach of $\varphi_{\pi1}(f_m)$ to $\varphi_{\pi2}(f_m)$ are connected with the above mentioned frequency effect.

Under increase of $f_m$ the capacitive part of impedance of SE-load circuit $1/2\pi f_m C_e$ value falls down and becomes less than the impedance of the load resistor $R_L$. So the common impedance of SE-load parallel circuit changes from resistive to capacitive one. At this the inequality $2\pi f_m R_L C_e << 1$ is changed to the opposite $2\pi f_m R_L C_e >> 1$. This transition in correspondence with (4) manifests itself in the change of the course of dependences $U_{\pi1}(f_m)$ and $\varphi_{\pi1}(f_m)$ and leads to their confluence with $U_{\pi2}(f_m)$ and $\varphi_{\pi2}(f_m)$. The evaluation of the frequency of this transition $f_{mt}$ was performed from $2\pi f_{mt}R_LC_e = 1$ with account of known $R_L$ and $C_e$ values. This estimation gives $f_{mt} \approx 330$ Hz for the free PZT ceramic SE and $f_{mt} \approx 6500$ Hz for the free PVDF film SE that is in agreement with data on the Figs 2a and 2b. It is obviously that the difference of the $f_{mt}$ values for the PZT ceramic and PVDF film samples is connected with differences in their dielectric constant values and their geometries.

*5.2. PZT-film sensitive elements on Si substrate*

The results obtained for PZT film based SE on Si substrate are presented in Fig. 3a and Fig. 3b. The influence of the substrate manifests itself by two factors: the first one is the essential suppression of the pyroelectric response value (substrate thermal capacity effect) and the second one is the change of the shape of $U_{\pi1,2}(f_m)$ and $\varphi_{\pi1,2}(f_m)$ dependences (substrate heat sink effect).

**1.** In fact the observed considerable difference in $U_\pi$ values of Pt-PZT-film – Si-substrate structure and Ag-PZT-bulk sample (the ratio of responses at the same low frequencies is $\approx 40$ for $U_{\pi2}$ values) is connected with the reasons of different physical nature. Firstly, it is the difference in absorptivity values of mirror-like Pt top electrode ($\alpha \approx 0.1$) for the Pt-PZT film-Si structure and gray ($\alpha \approx 0.5$) oxidized Ag electrode for the Ag-PZT plate; however, it gives only $\approx 5$ times of suppression.

Secondly, it is the intensive heat sink from PZT-film into Si-substrate. Indeed, the ratio of effusivities $b_t = c_t (a_t)^{1/2}$ for bulk Si and PZT-ceramic is $(b_t)_{Si}/(b_t)_{PZT} \sim 10$ and so Si-substrate is a good heat sink for PZT-ceramic. For investigated Si-substrate and PZT-film thickness the linear thermal capacities ratio is $(c_tL_S)_{Si}/(c_td)_{PZT} \approx 100$. So, expected "optical + thermal" suppression of response is $\approx 500$.

Taking into account the ratio $A_0/A_e$ in formulae (6), which is $\approx 2$ for PZT-film sample and is $\approx 30$ for PZT-ceramic bulk sample, we obtain the value of suppression $\approx 33$, which is near to the experimental ratio. On the other hand, it demonstrates a high pyroelectric quality of investigated PZT-film near to that of bulk PZT ceramics.

**2.** The behavior of pyroelectric response in Figs 3a and 3b in comparison with Fig. 2a demonstrates the existence of pronounced "substrate heat sink effects" for the investigated PZT-film – Si-substrate structures. Indeed, the presence of negative addition to $\varphi_{\pi1,2}$ which is about –45° at the low $f_m$, diffuse maximum of $U_{\pi1}(f_m)$, the break and only weak fold of $U_{\pi 2}(f_m)$, and also increase of $\varphi_{\pi 1,2}(f_m)$ at high $f_m$ are the evidence of the "substrate heat sink effects".

So the analysis of obtained dependences should be performed taking into account all aspects of substrate influence. The critical frequency region reflected the "substrate heat sink effects" is connected with the change of mode of the heat spreading, namely with transition from so-called "through substrate" mode through "inside substrate" mode to "under-electrode" mode. The characteristic modulation frequencies $f_{mc}$ of these transitions are determined by the ratios between the values of $\lambda_t$, effective electrode radius $r_e$ and substrate thickness $L_S$.

In the "under-electrode" mode $\lambda_t \ll r_e < L_S$ and the heated volume $V_t(f_m)$ of substrate decreases with increasing $f_m$ value mainly due to decrease of heat penetration depth $L_t \approx \lambda_t \propto f_m^{-1/2}$ along the substrate thickness. So, approximately $V_t(f_m) \propto f_m^{-1/2}$.

In the "inside substrate" mode $\lambda_t \approx r_e < L_S$ and the heated volume $V_t$ of substrate decreases with increasing $f_m$ value both due to decrease of heat penetration depth $L_t \approx \lambda_t$ along the substrate thickness and decrease of lateral heat spreading radius $R_{tS} = r_e + \lambda_t$ along the substrate surface. So, approximately $V_t(f_m) \propto f_m^{-3/2}$. The estimation of $f_{mc}^*$ from equality $\lambda_t = r_e$ with known electrode area $r_e = (A_e/\pi)^{1/2} \approx 0.5$ mm) and thermal diffusivity of crystalline Si gives $f_{mc}^* \approx 100$ Hz.

In the "through substrate" mode $\lambda_t \geq L_S$ and the heated volume of substrate decreases with increasing $f_m$ value only due to decrease of lateral spreading volume which is determined by square of radius of heat spreading $R_{tS}^2 = (r_e + \lambda_t)^2$. So, approximately $V_t(f_m) \propto f_m^{-1}$. The estimation of $f_{mc}^{**}$ from equality $\lambda_t = L_S$ with known thickness of Si-substrate and thermal diffusivity of Si gives $f_{mc}^{**} \approx 230$ Hz.

For investigated PZT on Si structures (Figs 3a and 3b) like for the free PZT ceramic SE (Fig. 4a) at low modulation frequencies both pyroelectric voltage and current modes are realized. At high $f_m$ the coincidence of the dependences $U_{\pi 1}(f_m)$ and $U_{\pi 2}(f_m)$ as well as the dependences $\varphi_{\pi 1}(f_m)$ and $\varphi_{\pi 2}(f_m)$ is observed. Thus, the frequency effect is clearly manifested. At that, there is the transition from pyroelectric current mode to pyroelectric voltage mode corresponding to increase of $\omega_m R_e C_e$ value and so, the only pyroelectric voltage mode for the PZT-film on the substrate is realized (Figs 3a and 3b). The evaluation of the frequency of this transition $f_{mt}$, when $U_{\pi 1}(f_m)$ approaches to $U_{\pi 2}(f_m)$ and the same behaviour is observed for $\varphi_{\pi 1}(f_m)$ and $\varphi_{\pi 2}(f_m)$), was performed with account of known $R_L$, $R_s$ and $C_s$ values. This estimation gives $f_{mt} \approx 170$ Hz for the PZT-film that corresponds to the data in Figs 3a and 3b.

Consideration of Figs 3a and 3b shows that the estimated critical frequencies $f_{mc}^*$, $f_{mc}^{**}$ and $f_{mt}$, are within the limits of observed peculiarities of $U_{\pi 1,2}(f_m)$ and $\varphi_{\pi 1,2}(f_m)$.

Our estimation of thermal diffusivity of investigated PZT film in correspondence with the relationship $d_{PZT} = \lambda_t$ under $d_{PZT} = 1.9$ µm and $f_{m2} \approx 40$ kHz gives $a_{tPZT} \approx 4.5 \cdot 10^{-7}$ m$^2$/s which is in the limits of literature values (see Table 1).

The shapes of $U_{\pi 1,2}(f_m)$ and $\varphi_{\pi 1,2}(f_m)$ are near identical for "+" and "-" poled SE (compare Figs 3a and 3b). Consecutive repolarization of SE gives 180°-addition to $\varphi_{\pi 1,2}$ which corresponds to the change of sign of pyroelectric reaction of SE and only insignificant variations of shape of $U_{\pi 1,2}(f_m)$ and $\varphi_{\pi 1,2}(f_m)$ dependences. This indicates the almost complete reorientation of polarization direction in inter-electrode space of PZT film. A small scatter of $U_{\pi 1,2}$ values can be explained by the difference in the degree of unipolarity of PZT film under the top Pt electrode and the bottom Pt/Ti one.

### 5.3. PZT-film sensitive elements on the por-Si intermediate layer

Results obtained for SE based on PZT film on intermediate porous Si layer on Si substrate are presented in Fig.4a and Fig 4b.

The comparison of the Figs 3a, 3b and Figs 4a, 4b shows that the inserting of buffer por-Si layer leads to the pronounced thermal decoupling of PZT film and Si substrate. The attenuation of heat sink influence of Si substrate is manifested in the changes of low and high frequency behavior of $U_{\pi 1,2}(f_m)$ and $\varphi_{\pi 1,2}(f_m)$ dependences. As the result of the increase of slope for $U_{\pi 1}(f_m)$ dependence and decrease of slope for $U_{\pi 2}(f_m)$ the pronounced maximum of $U_{\pi 1}(f_m)$ as well as the region of transition from near flat dependence to $1/f_m$ drop for $U_{\pi 2}(f_m)$ with flat region of $\varphi_{\pi 2}(f_m)$ are appeared.

Since such $1/f_m$ drop of $U_{\pi 2}(f_m)$ with flat region of $\varphi_{\pi 2}(f_m)$ is characteristic for a free sample (see Fig. 2a) it can be concluded that applied buffer por-Si layer at $f_m \geq 1$ kHz realizes a near complete thermal decoupling of PZT-film and Si-substrate.

From the extrapolated frequency of the break of $U_{\pi 2}(f_m)$ dependence $f_{mb} \approx 0.9$ kHz under condition that at $f_m = f_{mb}$ the temperature wave length $\lambda_t$ is equal to thickness $d_{por-Si}$ of por-Si layer it is possible to evaluate the diffusivity value $a_{t\,por-Si}$ of the por-Si layer. The estimation of $a_{t\,por-Si}$ from the equality $d_{por-Si} = \lambda_t$ for $d_{por-Si} = 10$ µm gives $a_{t\,por-Si} \approx 3 \cdot 10^{-7}$ m$^2$/s.

The obtained value of $a_{t\,por\text{-}Si}$ for por-Si with 65 % degree of porosity is in a good agreement with our previous estimation $(3\text{-}4)\cdot 10^{-7}$ m$^2$/s performed for por-Si with 60 % degree of porosity in the structures LiNbO$_3$-por-Si/Si [28].

The shapes of $U_{\pi 1,2}(f_m)$ and $\varphi_{\pi 1,2}(f_m)$ dependences are near identical for "+" and "-" poled SE (compare Figs 4a and 4b). The insignificant variations of shape of $U_{\pi 1,2}(f_m)$ dependences can be connected with a slight difference of poling state of PZT-film under the top Pt electrode and the bottom Pt/Ti one. The similar repoling reaction of PZT on SiO$_2$/Si (Figs. 3a, b) and PZT on por-Si/Si (Figs. 4a, b) structures under investigation indicates the similarity of the poling state in the inter-electrode space of the Pt-PZT film-Pt/Ti capacitor on the SiO$_2$/Si and on the por-Si/Si substrates.

As it is clearly seen from Figs 4a and 4b for investigated PZT on por-Si/Si structures under changing $f_m$ value the frequency effect is manifested by transition from the well-distinguished pyroelectric current and voltage modes to only one pyroelectric voltage mode. The evaluated transition frequency $f_{mt} \approx 900$ Hz for the PZT-film corresponds to the data in Figs 4a and 4b. Obtained for "+" and "-" poled PZT film the near equal values of $f_{mt}$ correspond to the similarity of "+" and "-" poling state of Pt-PZT film-Pt/Ti capacitor on the por-Si layer. A small difference in $U_{\pi 1,2}$ values after "+" and "-" poling can be connected with a small difference in poling state of PZT film under the top Pt electrode and the bottom Pt/Ti one.

*5.4. PVDF film sensitive elements on the Si substrate.*

The results obtained for the free SE based on PVDF film on Si substrate are presented in Fig. 5a and Fig. 5b.

The comparison of the obtained results for PVDF on Si and for PZT on por-Si/Si structures shows the similar behavior of $U_{\pi 1,2}(f_m)$ and $\varphi_{\pi 1,2}(f_m)$ dependences of structures under investigation (see Fig. 4a, b and Fig. 5a, b). This fact can be understood taking into account the lower values of thermal conductivity and diffusivity of PVDF comparatively with the same of por-Si. Indeed, the estimations of the values of linear thermal resistance $r_{th} = (1/k_t)d$ and time of the heat diffusion $\tau_t = d^{\,2}/2a_t$ of applied por-Si and PVDF layers gives $r_{th\,por-Si} \approx 3.3\cdot 10^{-5}$ m$^2$K/W and $\tau_{t\,por\text{-}Si} \approx \tau_{t\,PVDF} \approx 1.7\cdot 10^{-4}$ s. So, if for PZT on por-Si/Si structure thermal decoupling at frequencies $f_m \geq 1$ kHz is realized by por-Si layer, for the investigated PVDF on Si structure an effective thermal self-decoupling by a near substrate part of PVDF layer itself, which is thermally not excited, takes place.

The estimation of diffusivity $a_{t\,PVDF} = \pi \cdot f_{mb} d_{PVDF}^2$ for our samples of PVDF film from the frequency of break of $U_{\pi2}(f_m)$ dependence $f_{mb} \approx 1.2$ kHz gives $a_{t\,PVDF} \approx 7 \cdot 10^{-8}$ m$^2$/s which is rather close to known data for PVDF (see Table 1). This proximity also shows the validity of our estimation of thermal diffusivity of the investigated PZT film and por-Si layer.

The shapes of $U_{\pi1,2}(f_m)$ and $\varphi_{\pi1,2}(f_m)$ dependences are rather similar for "+" and "-" poled PVDF film (compare Figs 5a and 5b). The slight variations of shape of $U_{\pi1,2}(f_m)$ dependences reflect most probably the slight difference in "+" and "-" poling state of Al-PVDF film-Ti on Si structure under the top Al electrode and the bottom Ti one.

As it is seen from Figs 5a and 5b for investigated PVDF on Si structures under changing $f_m$ value the frequency effect is manifested. The evaluated transition frequency $f_{mt} \approx$ 3.1 kHz for the PVDF-film corresponds to the obtained data. Near equal values of $f_{mt}$ for "+" and "-" poled PVDF film correspond to the similarity of "+" and "-" poling state of Al-PVDF film -Ti capacitor on Si.

## 6. Discussion

**1.** The comparison of the results of the experiment (Figs. 2 – 5) and theoretical consideration (Fig. 1) shows a good agreement of model and observed behaviour of $U_{\pi1,2}(f_m)$ and $\varphi_{\pi1,2}(f_m)$ in the whole frequency range of performed experiment.

The results obtained for PZT on Si and PZT on por-Si/Si structure show that a porous Si layer is a suitable material for effective thermal decoupling of SE and heat removing substrate. It gives possibility to increase significantly pyroelectric response value and approach it to that of characteristic for a free SE using Si derived material on Si-substrate. Therefore, for obtaining a complete thermal decoupling of SE and substrate in the structure of PZT on por-Si/Si at the frequencies $f_m \geq 1$ kHz is enough to introduce of buffer por–Si layer of thickness 10 μm.

Undoubtedly, application of buffer layer from por-Si is also effective for PVDF on por-Si/Si structures. At the same time for PVDF on Si structures there is a possibility of obtaining complete thermal decoupling of SE and substrate at frequency $f_m \geq 1$ kHz by using PVDF layer of thickness more than 4.5 μm.

**2.** The comparison of $U_\pi(\omega_m)$ and $\varphi_\pi(\omega_m)$ dependences obtained for structures of PZT on por-Si/Si and PVDF on Si shows the possibility of evaluation of the main thermophysical parameters of buffer layer material $k_t$, $a_t$, and $c_t$ from the following expression

$$k_{t\,\text{por-Si}} = k_{t\,\text{PVDF}}(d_{\text{por-Si}}/d_{\text{PVDF}})(r_{th\,\text{PVDF}}/r_{th\,\text{por-Si}}),$$

$$a_{t\,\text{por-Si}} = a_{t\,\text{PVDF}}(d_{\text{por-Si}}/d_{\text{PVDF}})^2(f_{mb\,\text{por-Si}}/f_{mb\,\text{PVDF}}),$$

$$c_{t\,\text{por-Si}} = k_{t\,\text{por-Si}}/a_{t\,\text{por-Si}},$$

obtained by means of comparison the values of linear thermal resistance $r_{th}$ and time of the heat diffusion $\tau_t$.

**3.** For designing pyroelectric detectors with high sensitivity a good thermal contact with heat removing substrate is non-desirable. Thermal decoupling of SE and substrate can be achieved by the following ways:

- Increase of modulation frequency up to values more than $f_m = a_t/\pi d^2$, where $a_t$ and $d$ are the thermal diffusivity and thickness of SE material (frequency decoupling). This solution is easy to obtain by apparatus way but it is non-effective for SE material with large value of $a_t/d$ ratio and can be connected with decrease of signal-to-noise ratio at high $f_m$.

- Increase of SE thickness up to values more than $d_{\min} = (a_t/\pi f_{\min})^{1/2}$, where $f_{\min}$ is the minimal operation modulation frequency (thickness decoupling). This solution is achieved by technological way and is effective at small values of $a_t/f_{\min}$ (for material of SE with small $a_t$ value and under operation at high frequencies $f_m$) but it is connected with decreasing electrical capacity of SE that sometimes complicates electric matching.

- Insertion of thermal isolating interlayer between SE and substrate (thermal buffer) with thickness exceeding $d_{hd} = (a_{thd}/\pi f_{\min})^{1/2}$, where $a_{thd}$ is the thermal diffusivity of interlayer (material decoupling). This solution is more complicated technologically than those above mentioned but is more effective for SE materials in wide range of values of thermal diffusivity and thickness.

## 7. Conclusion remarks

Polar PVDF films are suitable pyroelectric material for pyroelectric detectors with SE on Si-substrate. Together with high enough value of pyroelectric figure of merit $M_2$, mechanical flexibility, chemical resistance and low cost of fabrication PVDF films with certain thickness can provide thermal self-decoupling of SE and substrate due to relatively low value of thermal diffusivity. Therefore, despite low value of pyroelectric figure of merit $M_1$ and not high temperature of reversible conservation of pyroactivity that rather complicates the technological procedure on the stage of soldering, they can compete with PZT films.

Polar PZT films are compatible with integrated Si technology and have a high temperature of reversible conservation of pyroactivity with a high value of pyroelectric figure of merit of $M_1$ and a high enough value of $M_2$. At the same time PZT films possess a high temperature of formation of polar phase and comparatively high thermal diffusivity. The first factor leads to difficulties connected with stability of semiconductor elements, the second one demands insertion of thermal isolating layer for thermal decoupling of SE and substrate.

Comparatively high $M_2$ values of polar PVDF films leads to more effective operation in the pyroelectric voltage mode, and comparatively high $M_1$ values of polar PZT films leads to more effective operation in the pyroelectric current mode. Using the known expressions for $U_\pi$ value and for thermal noise voltage [3] it can be shown that unlike its operation in the pyroelectric voltage mode when the noise equivalent power (NEP) is $P_{\mathrm{nv}} = (c_t/\alpha\gamma)\sqrt{4kT\,d\,A\sigma}$, in the pyroelectric current mode $P_{\mathrm{ni}} = (c_t/\alpha\gamma d)\sqrt{4kT/R_L}$. Thus these two operation modes display the different relationships of both the NEP and normalized detectivity D* = $A^{1/2}/P_n$ with the electrical and geometric parameters of the pyroelectric element and the external circuit parameters. In particular, the influence of variation of the thickness d of SE on NEP and D* in the pyroelectric current mode is more efficient than in the pyroelectric voltage mode.

The higher σ values of PZT films comparatively to those of PVDF films and the possibility of controlling σ by doping makes it unnecessary to use a leakage resistor in the input circuit of the matching stage in the pyroelectric voltage mode which is essential for design of integrated pyroelectric micro-devices.

Porous silicon is Si-derived material with comparatively low thermal diffusivity. For por-Si there is a possibility of high-speed formation of layer with desirable values of $a_t$ and $\tau_{th}$ changing its degree of porosity and thickness directly on the surface of Si-substrate. Due to this fact por-Si is a highly suitable material for application as a thermal decoupling layer between SE and Si-substrate of Si-integrated pyroelectric detectors.

## 8. Acknowledgement

Authors gratefully acknowledge Ministry of Science of France and the University of Valenciennes for financial support and also Yann Henrion for PVDF sample providing and elaboration.

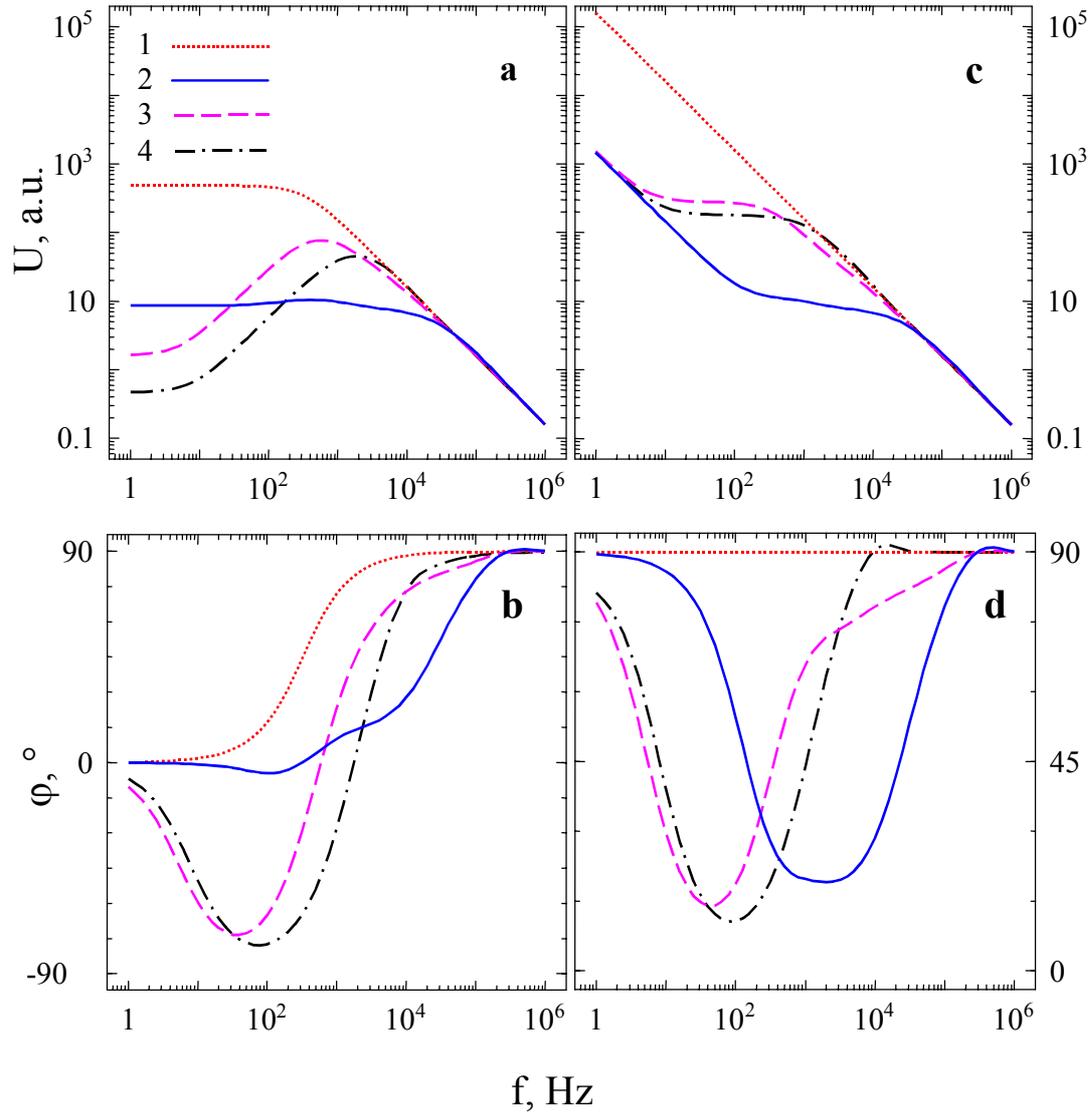

FIGURE. 1. The calculated amplitude-to-frequency $U_{\pi1,2}(f_m)$ and phase-to-frequency $\varphi_{\pi1,2}(f_m)$

dependences of pyroelectric response

in pyroelectric current mode: a – $U_{\pi1}(f_m)$, b - $\varphi_{\pi1}(f_m)$, and

in pyroelectric voltage mode: c – $U_{\pi2}(f_m)$, d - $\varphi_{\pi2}(f_m)$

for SE systems based on :

- free Ag-PZT-Ag and Al-PVDF-Al (curves 1),
- Pt-PZT-Pt/Ti/SiO$_2$/Si (curves 2),
- Pt-PZT-Pt/Ti/porSi/SiO$_2$/Si (curves 3),
- Al/Ti-PVDF-Ti/SiO$_2$/Si (curves 4).

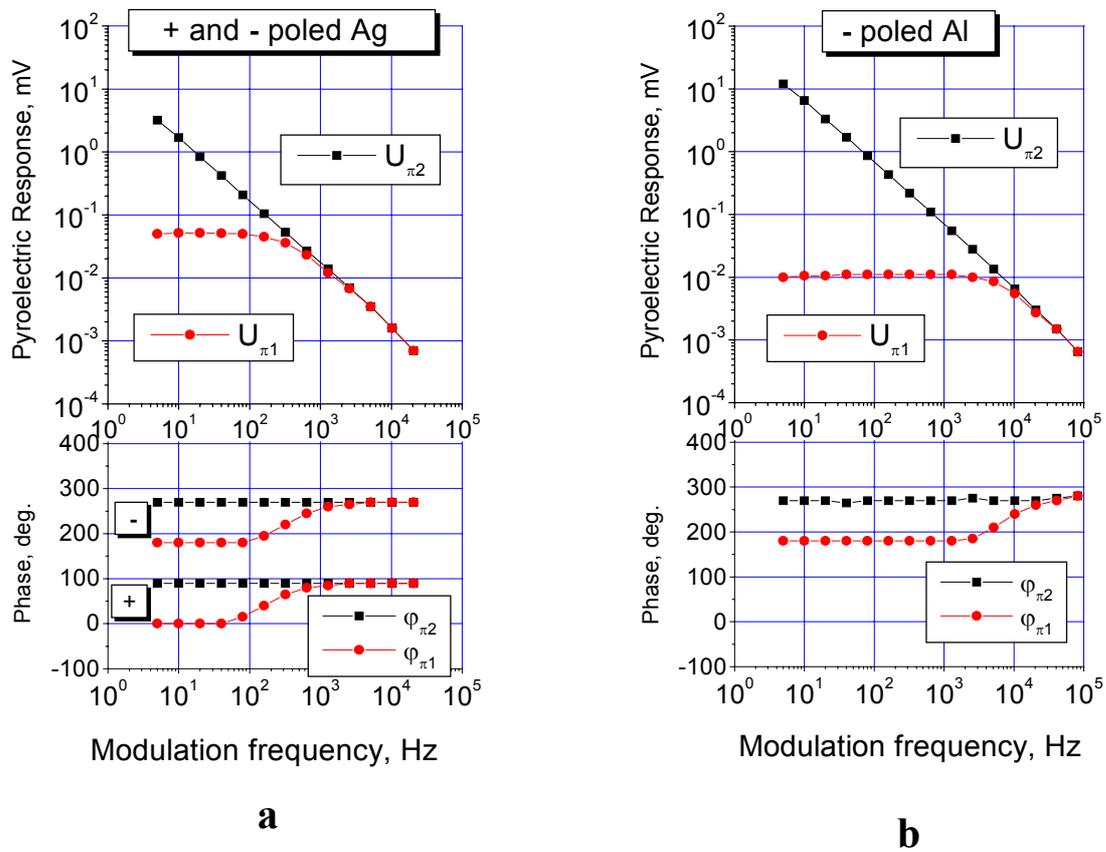

FIGURE. 2. The dependences of $U_{\pi1,2}(f_m)$, and $\varphi_{\pi1,2}(f_m)$ of:

a - free Ag-PZT-Ag, 200 μm thick PZT, Φ6 mm Ag-electrodes,

D. C. poled; C = 4850 pF, G = 0,8 mS;

b - free Al-PVDF-Al, 7 μm thick PVDF, 7x2,5 mm Ag-electrodes,

Spin coated, Corona charged; C = 243 pF, G = 0,03 mS.

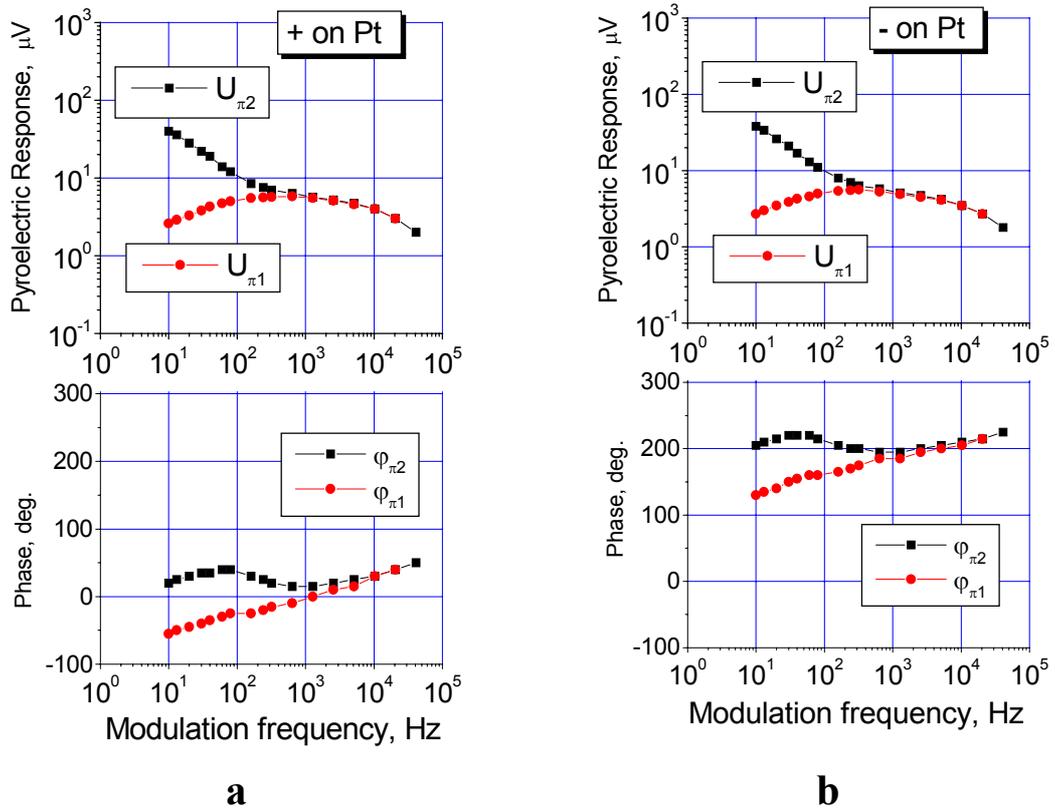

FIGURE. 3. The dependences of $U_{\pi1,2}(f)$, and $\varphi_{\pi1,2}(f)$ of Pt-PZT-Pt/Ti/SiO$_2$/Si (2 mm$^2$ Pt-electrode, 1,9 μm thick PZT-film on 0,35 mm Si-substrate)

a - +35 V, 20 °C, 5 min DC poled; C = 9420 pF, G = 2 mS;

b - -35 V, 20 °C, 5 min DC poled; C = 9510 pF, G = 1,1 mS.

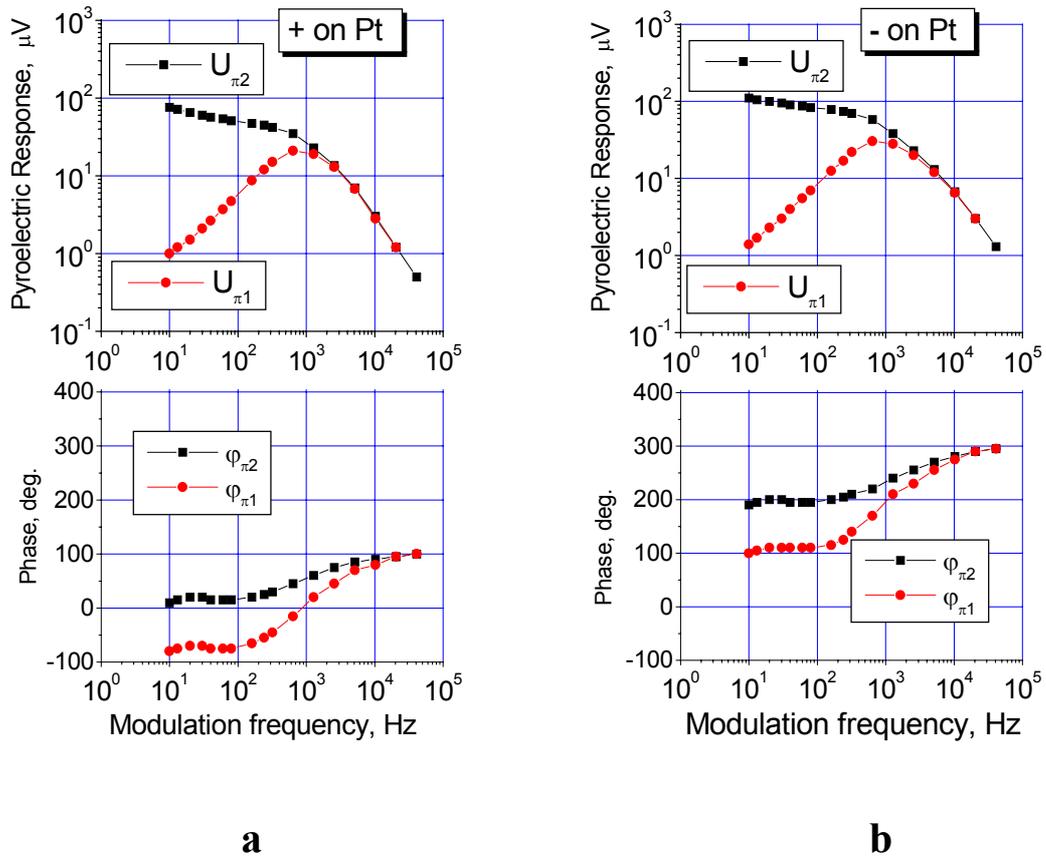

        **a**                 **b**

FIGURE. 4. The dependences of $U_{\pi 1,2}(f)$, and $\varphi_{\pi 1,2}(f)$ of Pt-PZT-Pt/Ti/porSi/Si (1x1 mm Pt-electrode, 1,4 μm thick PZT-film on 10 μm por-Si interlayer on 0,35 mm Si-substrate)

a - +24 V, 20 °C, 5 min DC poled; C = 1720 pF, G = 0,19 mS;

b - -24 V, 20 °C, 5 min DC poled; C = 1770 pF, G = 0,2 mS.

2424

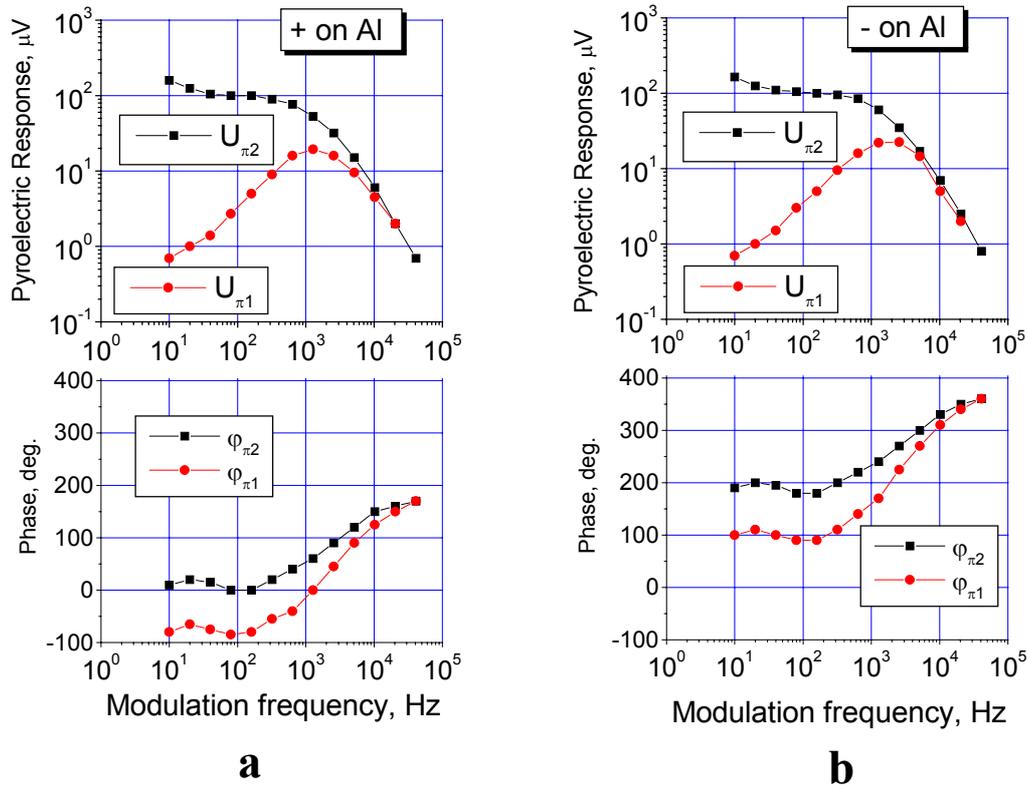

FIGURE. 5. The dependences of $U_{\pi1,2}(f_m)$, and $\varphi_{\pi1,2}(f_m)$ of Al-PVDF-Ti/Si (5x5 mm Al-electrode, 4,3 μm thick PVDF-film on 0,6 mm Si-substrate)

a - +200 V, 60 °C, 2 min DC poled; C = 512 pF, G = 0,07 mS

b - -200 V, 60 °C, 2 min DC poled; C = 502 pF, G = 0,06 mS